\documentclass{moriond}

\newcommand{\Photo}{\includegraphics[height=35mm]{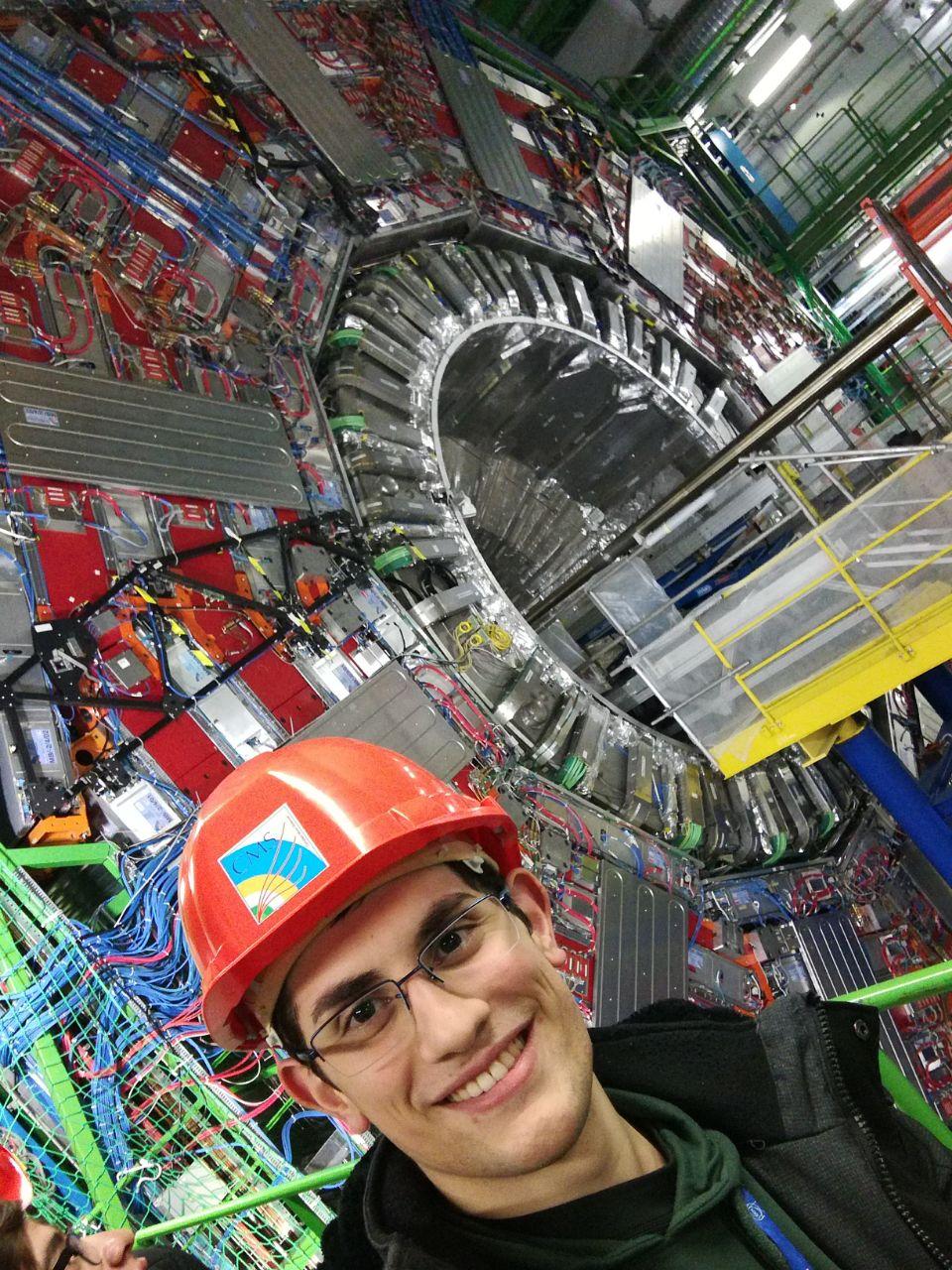}}

\newcommand{\gF}{\ensuremath{g_{\mathrm{F}}}}
\newcommand{\gH}{\ensuremath{g_{\mathrm{H}}}}
\newcommand{\gql}{\ensuremath{g_{\mathrm{q}_{12}}}}
\newcommand{\gqh}{\ensuremath{g_{\mathrm{q}_{3}}}}
\newcommand{\gll}{\ensuremath{g_{\mathrm{\ell}_{12}}}}
\newcommand{\glh}{\ensuremath{g_{\mathrm{\ell}_{3}}}}

\begin{document}
\vspace*{4cm}
\title{Searches for New Physics at High Object Masses with CMS}

\author{Andrea Malara}

\address{On behalf of the CMS Collaboration~\footnote{Copyright 2026 CERN for the benefit of the CMS Collaboration. Reproduction of this article or parts of it is allowed as specified in the CC-BY-4.0 license.}}

\maketitle\abstracts{
Searches at high object masses probe both resonant production of new particles and nonresonant
distortions of Standard Model spectra. This contribution follows the material presented in
the Moriond Electroweak 2026 talk and summarizes recent CMS results in this regime:
the Run~2 combination of heavy vector boson searches,
the Run~3 search for $W^\prime \to \ell \nu$, the Run~2 dijet angular analysis,
and searches for pair-produced dijet resonances in inclusive and $b$-tagged final states.
No significant deviation from the SM expectation is observed,
and the new results extend the sensitivity of CMS to multi-TeV scales in several benchmark scenarios.
}

\section{Introduction}
To date, the Standard Model (SM) provides a remarkably precise description of particle interactions,
which has been validated by decades of experiments, including measurements at the TeV scale at the LHC.

Despite its success, the SM leaves several fundamental questions unanswered,
including the nature of dark matter, the origin of the matter-antimatter asymmetry,
and the stability of the electroweak scale.
Many extensions of the SM attempt to address these puzzles.
Such theories often predict additional bosons, dark matter candidates, or extra dimensions
that could manifest at the LHC.
Because the parameters of these models are generally unknown,
their signatures can appear in a variety of final states involving jets,
leptons, or missing transverse momentum, depending on their couplings to SM particles.

The CMS experiment~\cite{CMS:2008xjf,Hayrapetyan_2024} explores this broad landscape with analyses targeting different final states.
The results discussed in these proceedings span both Run~2 and Run~3 data and
illustrate the complementarity between fermionic and bosonic final states,
as well as between resonant and nonresonant search strategies.

% Proably not fitting the text.
% New physics at the LHC can appear in two complementary ways: as a resonant peak associated with the production of a new particle,
% or as a broad distortion in the tails of kinematic distributions when the new scale is too high to produce a narrow resonance directly.

\section{Heavy vector boson searches}
\subsection{Run~2 legacy combination}

A combination of 16 searches for heavy spin-1 resonances covering many final states was performed,
using the Run~2 data set of 138~fb$^{-1}$ at $\sqrt{s}=13$~TeV~\cite{CMS:2026fjw}.
The combination includes final states with pairs of W, Z, or Higgs bosons,
as well as quark pairs ($qq$, $bb$, $tt$, $tb$) or lepton pairs ($\ell\ell$, $\ell\nu$), with $\ell$ = e, $\mu$, $\tau$.
The results are interpreted in the Heavy Vector Triplet (HVT) framework,
a simplified model for spin-1 resonances in which the couplings to fermions and bosons are
controlled by a small set of parameters.
The fermionic and bosonic final states are complementary because
they probe different couplings and regions of parameter space.

One of the main motivations for this combination was to clarify several local fluctuations
at the 2 and 3$\sigma$ level reported by individual searches.
In the combined result, these tensions are reduced and no significant deviation from the SM expectation is observed.
The analysis excludes heavy vector resonances below 5.5~TeV in the weakly coupled scenario,
below 4.8~TeV in the strongly coupled scenario, and up to 2~TeV for vector boson fusion production.

\begin{figure}[t]
\centering
\includegraphics[width=0.49\linewidth]{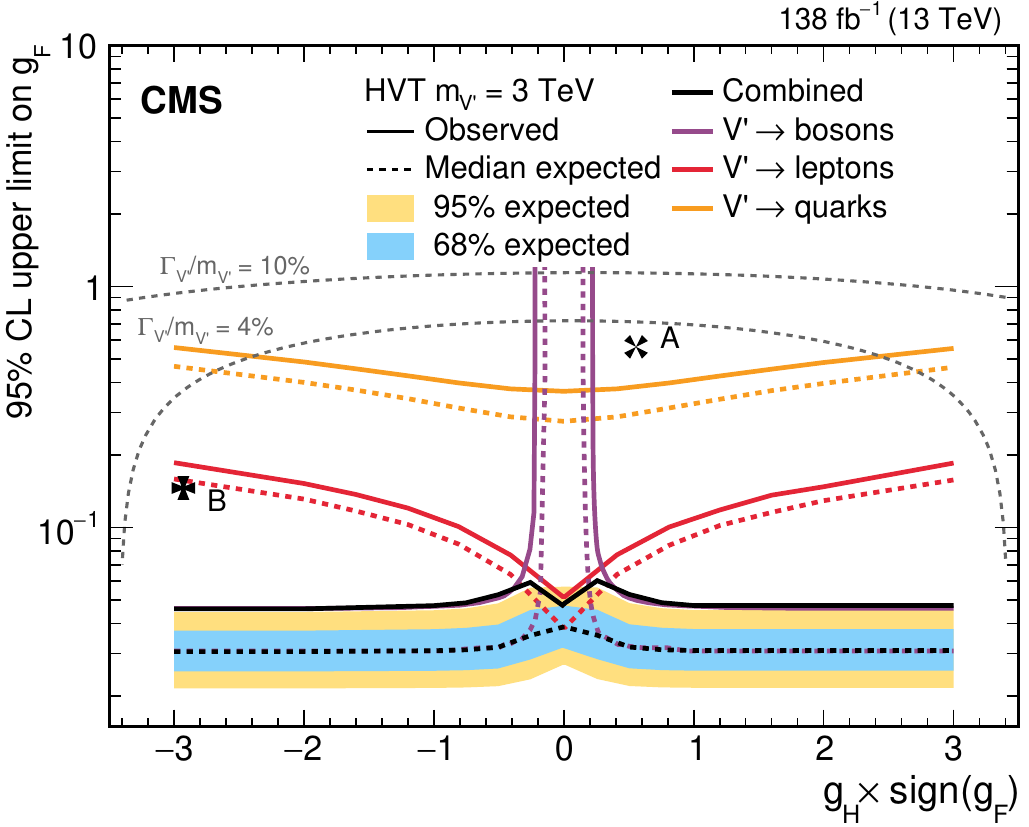}
\includegraphics[width=0.49\linewidth]{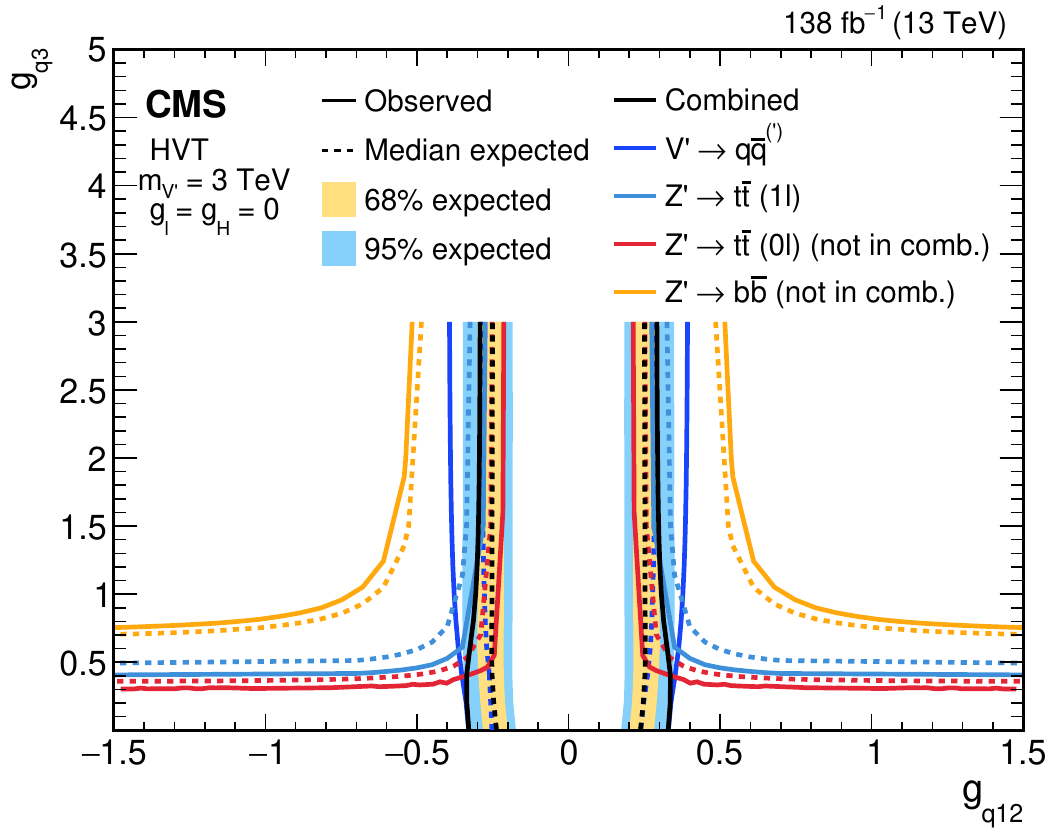}
\caption{Expected and observed 95\% CL upper limits on the coupling parameter \gF~as functions of \gH~(left)
and \gql~as functions of \gqh~(right) for resonance masses of 3~TeV.}
\label{fig:combination}
\end{figure}

Thanks to the complementarity of bosonic and fermionic channels,
a significant portion of the phase space for fermionic (\gF) and bosonic (\gH) couplings has been excluded.
The expected and observed 95\% confidence level (CL) upper limits on the coupling parameters are shown in Figure~\ref{fig:combination}.

Scenarios beyond the assumption of fermion universality are also explored,
and exclusion limits are evaluated in the \gqh-\gql~plane, with \gll, \glh, and \gH~set to zero.
This probes the possibility of nonuniversal couplings between light and third-generation quarks.

Additional scans are performed in the \gqh-\glh~and \gqh-\gH~planes.
The resulting parameter regions are fully excluded at 95\% CL,
driven by the combined sensitivity of the quark, lepton, and bosonic categories.
The complementarity of the individual channels allows the complete exclusion of these parameter spaces,
which could not be achieved by any single channel alone.

\subsection{Run~3 search for \texorpdfstring{$W^\prime \to \ell \nu$}{W' to l nu}}

The search for a charged heavy vector boson decaying to an electron or muon
and missing transverse momentum uses 62~fb$^{-1}$ of Run~3 data collected in 2022 and 2023
at $\sqrt{s}=13.6$~TeV~\cite{CMS-PAS-EXO-24-021}.
This channel remains one of the most sensitive probes of heavy vector bosons
with a clean experimental signature, consisting of one isolated high-$p_{\mathrm{T}}$ lepton
and large missing transverse momentum. The discriminant is the transverse mass.
The dominant background is SM $W$+jets production,
modeled with NNLO QCD and NLO electroweak corrections.
At high values of the transverse mass, the leading uncertainties arise from the lepton and jet momentum scale and resolution,
as well as from the SM $W$+jets background cross section prediction.

\begin{figure}[t]
\centering
\includegraphics[width=0.49\linewidth]{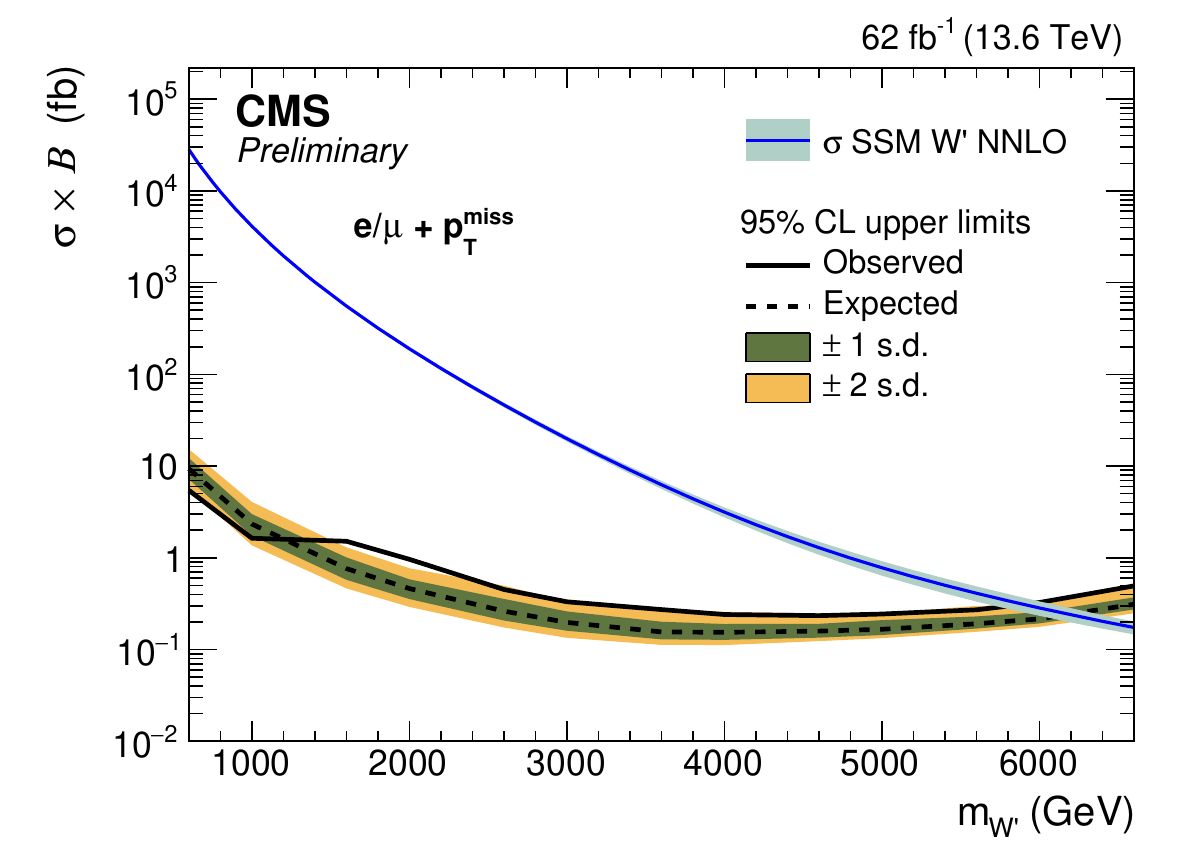}
\includegraphics[width=0.49\linewidth]{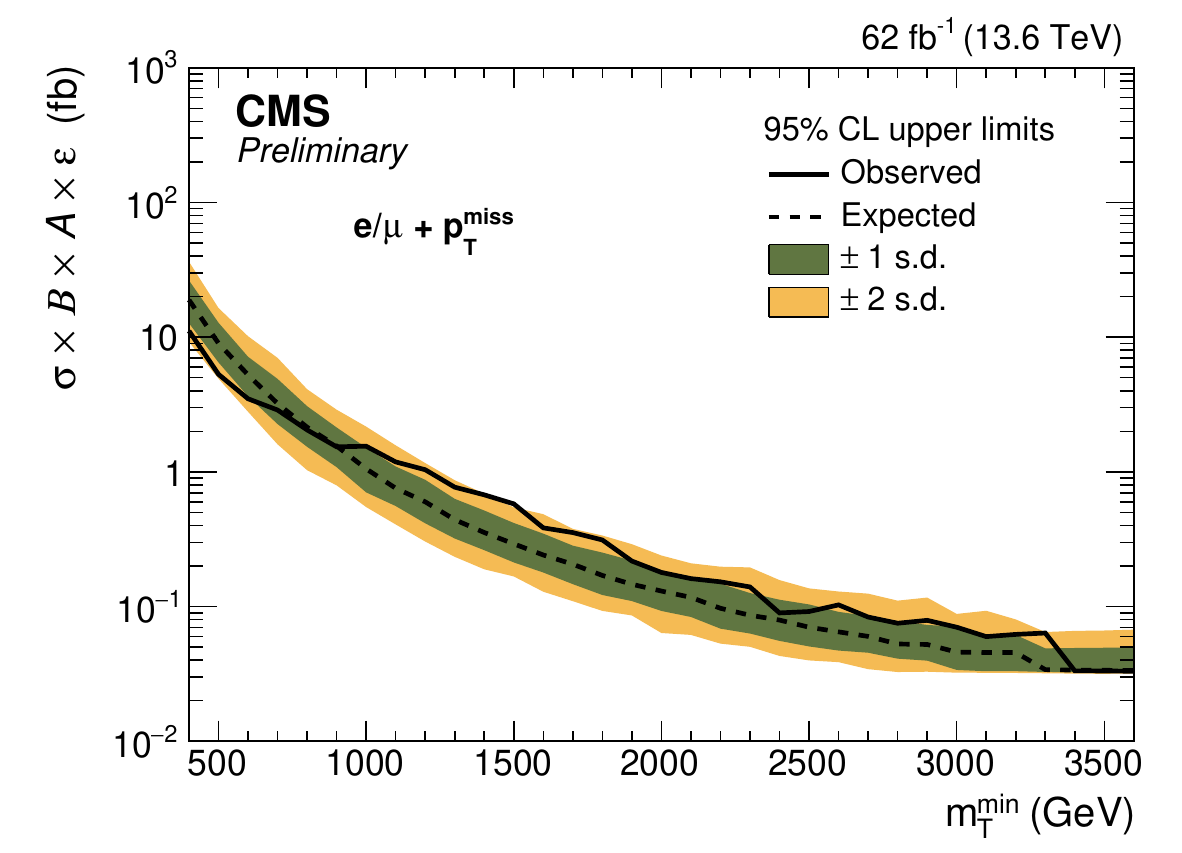}
\caption{Expected and observed 95\% CL upper limits for the $W^\prime \to \ell \nu$ search
in the sequential SM benchmark (left) and in the model-independent interpretation (right).}
\label{fig:wprime}
\end{figure}

No evidence for physics beyond the SM is found.
In the sequential SM benchmark, 95\% CL exclusions are set at 5.7~TeV in the electron channel,
5.6~TeV in the muon channel, and 5.9~TeV in the combination of both channels,
with an expected combined sensitivity of 6.2~TeV.
The analysis also reports model-independent limits based on event counts
above a minimum transverse-mass threshold,
allowing reinterpretations for scenarios with large widths or interference effects.
The upper limits for a $W^\prime$ boson with SM-like couplings are shown in Figure~\ref{fig:wprime}.
Compared with the full Run~2 CMS result, the improved sensitivity
is driven by the higher collision energy and by the lower single-electron trigger threshold.

\section{High-mass dijet signatures}
\subsection{Dijet angular analysis}

A complementary approach to the search for new physics
consists of probing nonresonant distortions in the tails of SM predictions.
An example is the dijet angular analysis~\cite{CMS:2026ecv},
which is sensitive to changes in the shape of the scattering angle $\theta^*$ of the dijet system
through the variable
\begin{equation}
\chi_{\mathrm{dijet}} \sim \frac{1+\cos(\theta^*)}{1-\cos(\theta^*)}\;,
\end{equation}
which exhibits a mild dependence on theoretical modelling due to parton distribution functions
and on the strong coupling constant.
The analysis uses the full Run~2 data set of 138~fb$^{-1}$ at $\sqrt{s}=13$~TeV.
Detector-level normalized distributions are used to search for beyond-the-SM effects,
through a comparison with next-to-next-to-leading-order QCD predictions including
next-to-leading-order electroweak corrections, smeared to account for the detector response.

\begin{figure}[t]
\centering
\includegraphics[width=0.49\linewidth]{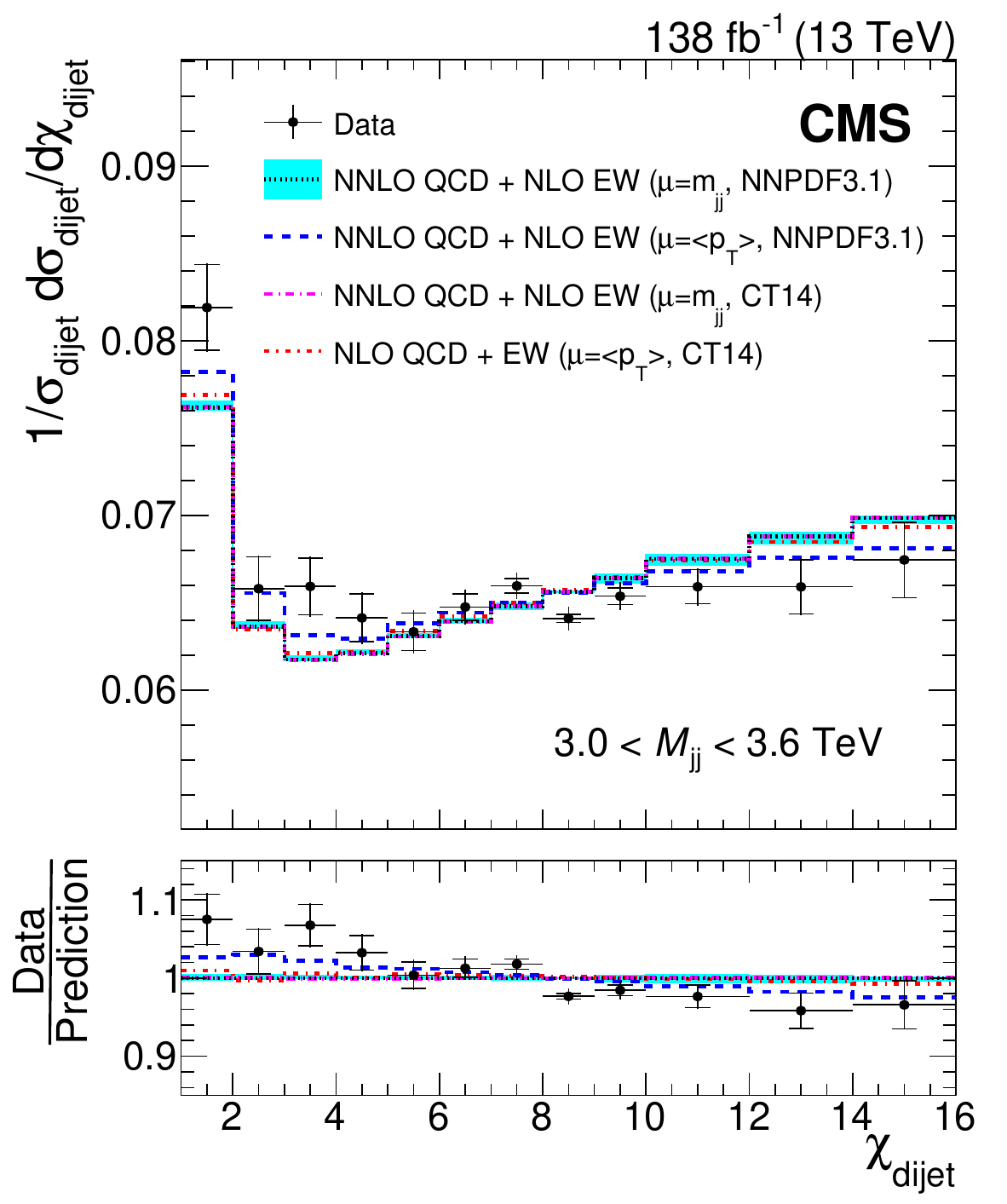}
\includegraphics[width=0.49\linewidth]{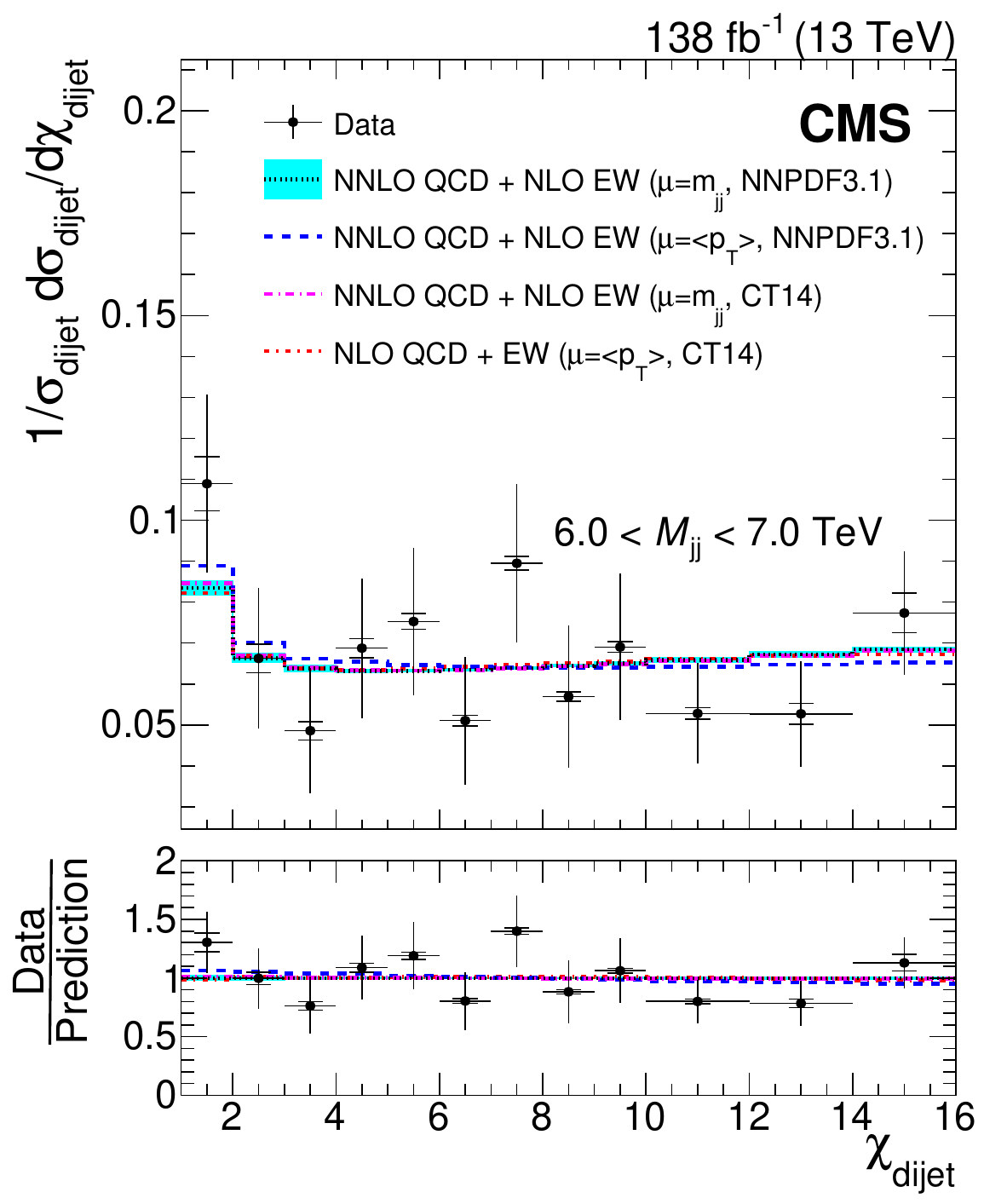}
\caption{Representative normalized $\chi_{\mathrm{dijet}}$ distributions in data compared with the NNLO QCD + NLO EWK prediction.
Benchmark beyond-the-SM expectations are overlaid.}
\label{fig:angular}
\end{figure}

The data are generally well described by the predictions, as shown for example in Figure~\ref{fig:angular}.
A mild shape difference is observed for dijet masses from 2.4 to 4.8~TeV and above 6~TeV
with respect to the default NNLO prediction, while an alternative scale choice gives good agreement.
No evidence for beyond-the-SM physics is found.

The results are interpreted by setting exclusion limits at 95\% CL on the new-physics scale
in a broad set of beyond-the-SM scenarios,
covering quark contact interactions, virtual graviton exchange,
quantum black holes, dark-matter mediators, axion-like particles,
and anomalous gluon couplings in effective field theory.
While the precise values depend on the model, they reach scales of several tens of TeV,
illustrating the sensitivity of the analysis to high-mass effects well beyond the direct energy reach of the LHC.

\subsection{Pair-produced dijet resonances in Run~3}

The CMS Collaboration has updated the search for resonant production of
pairs of dijet resonances using 90~fb$^{-1}$ of 2024 data at $\sqrt{s}=13.6$~TeV~\cite{CMS-PAS-EXO-25-004}.
The analysis targets events with at least four resolved jets and invariant mass $m_{4j}>1.6$~TeV.
To avoid sculpting in the $m_{4j}-m_{2j}$ plane, where $m_{2j}$ is the average mass of the two dijet systems,
the search is performed in thirteen bins of $\alpha = m_{2j}/m_{4j}$ and uses smooth empirical functions
to model the falling multijet background in each $m_{4j}$ spectrum.

\begin{figure}[t]
\centering
\includegraphics[width=0.49\linewidth]{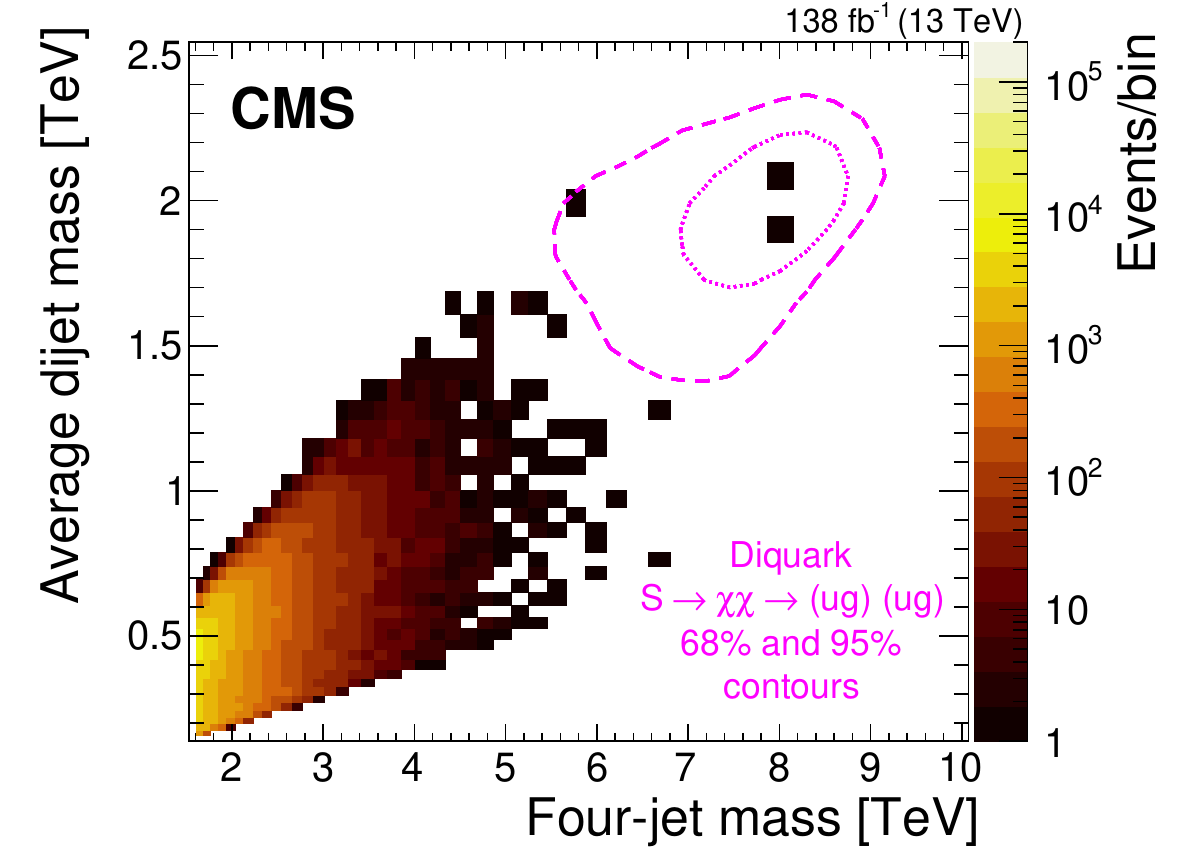}
\includegraphics[width=0.49\linewidth]{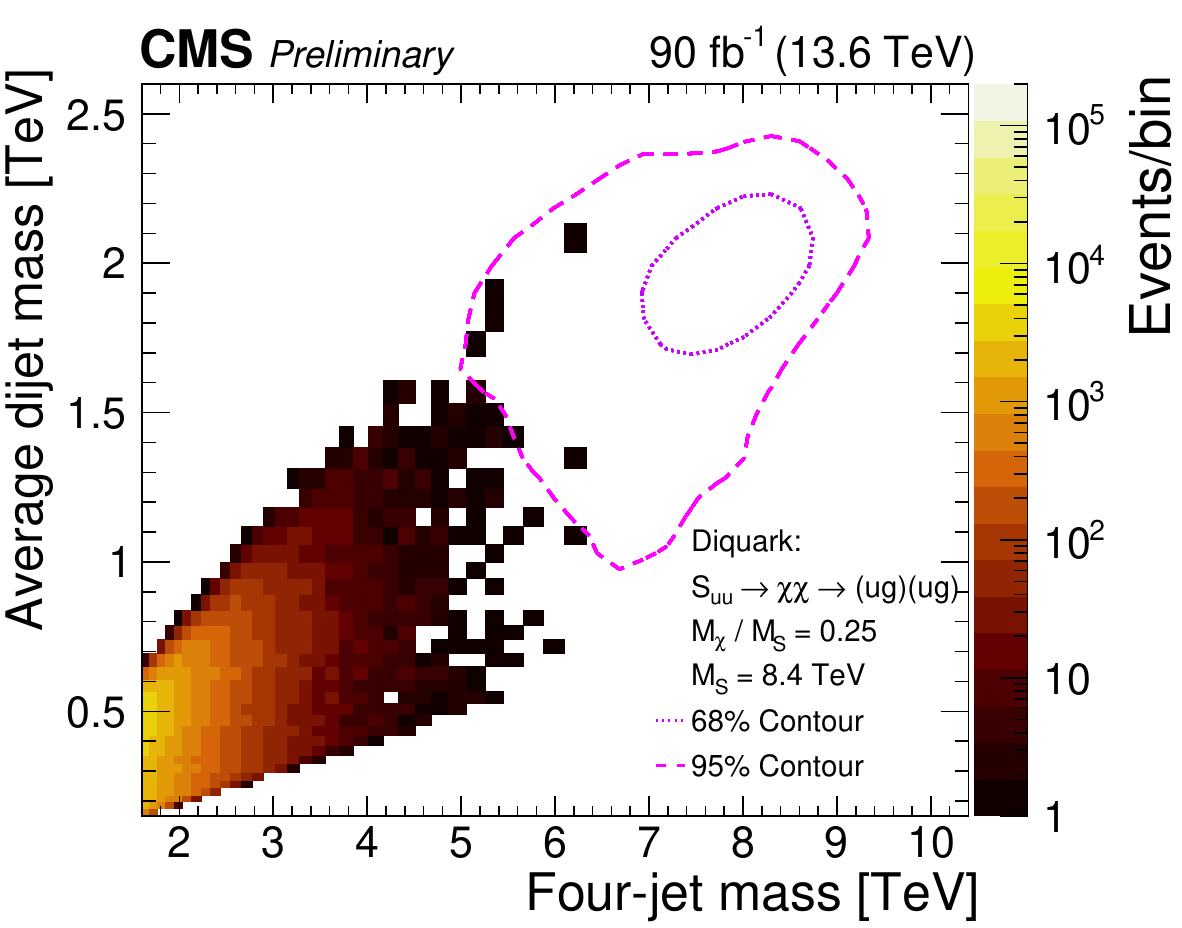}
\caption{Observed events in the $m_{4j}-m_{2j}$ plane for the 
for the previous Run~2 analysis with 138~fb$^{-1}$ at 13~TeV (left)
and the new Run~3 search with 90~fb$^{-1}$ at 13.6~TeV (right).
The dotted and dashed curves show the 68\% and 95\% probability contours, respectively,
from a signal simulation of a diquark with a mass of 8.4 TeV,
decaying to a pair of vector-like quarks, each with a mass of 2.1 TeV.}
\label{fig:pairdijet}
\end{figure}

The event selection and jet-pairing strategy are kept identical to the previous Run~2 analysis,
which had reported two events near $m_{4j}\approx 8$~TeV and $m_{2j}\approx 2$~TeV,
corresponding to a local (global) significance of 3.9 (1.6) standard deviations.
With the 2024 sample, no event is observed in the 8~TeV region.
A high-mass event with $m_{4j}\approx 6$~TeV and $m_{2j}\approx 2$~TeV is observed,
so the new data do not confirm the Run~2 excess.
The largest fluctuation instead appears at $m_{4j}=4.7$~TeV and $m_{2j}=0.89$~TeV,
with a local (global) significance of 3.3 (1.0) standard deviations.
Figure~\ref{fig:pairdijet} compares the Run~2 and Run~3 event distributions
in the $m_{4j}-m_{2j}$ plane.

In the absence of any deviation from the expectation, upper limits are set on
the production cross section times branching fraction and acceptance for four-jet resonance masses between 2 and 9~TeV.
In the benchmark scalar-diquark model, which predicts a decay to a pair of vector-like quarks, each of which decays to a quark--gluon pair,
masses below 6.3 and 8~TeV are excluded, depending on the coupling parameters.

\subsection{Pair-produced dijet resonances with \texorpdfstring{$b$}{b} jets}

A similar Run~2 search targets pair-produced dijet resonances in final states with one $b$ jet
and one light-flavor jet per dijet candidate, using the full 138~fb$^{-1}$ data set at $\sqrt{s}=13$~TeV~\cite{CMS-PAS-EXO-24-039}.
Results are presented separately for nonresonant production,
interpreted in an R-parity-violating (RPV) top-squark model,
and resonant production in the diquark-to-diquarks model.

\begin{figure}[h]
\centering
\includegraphics[width=0.49\linewidth]{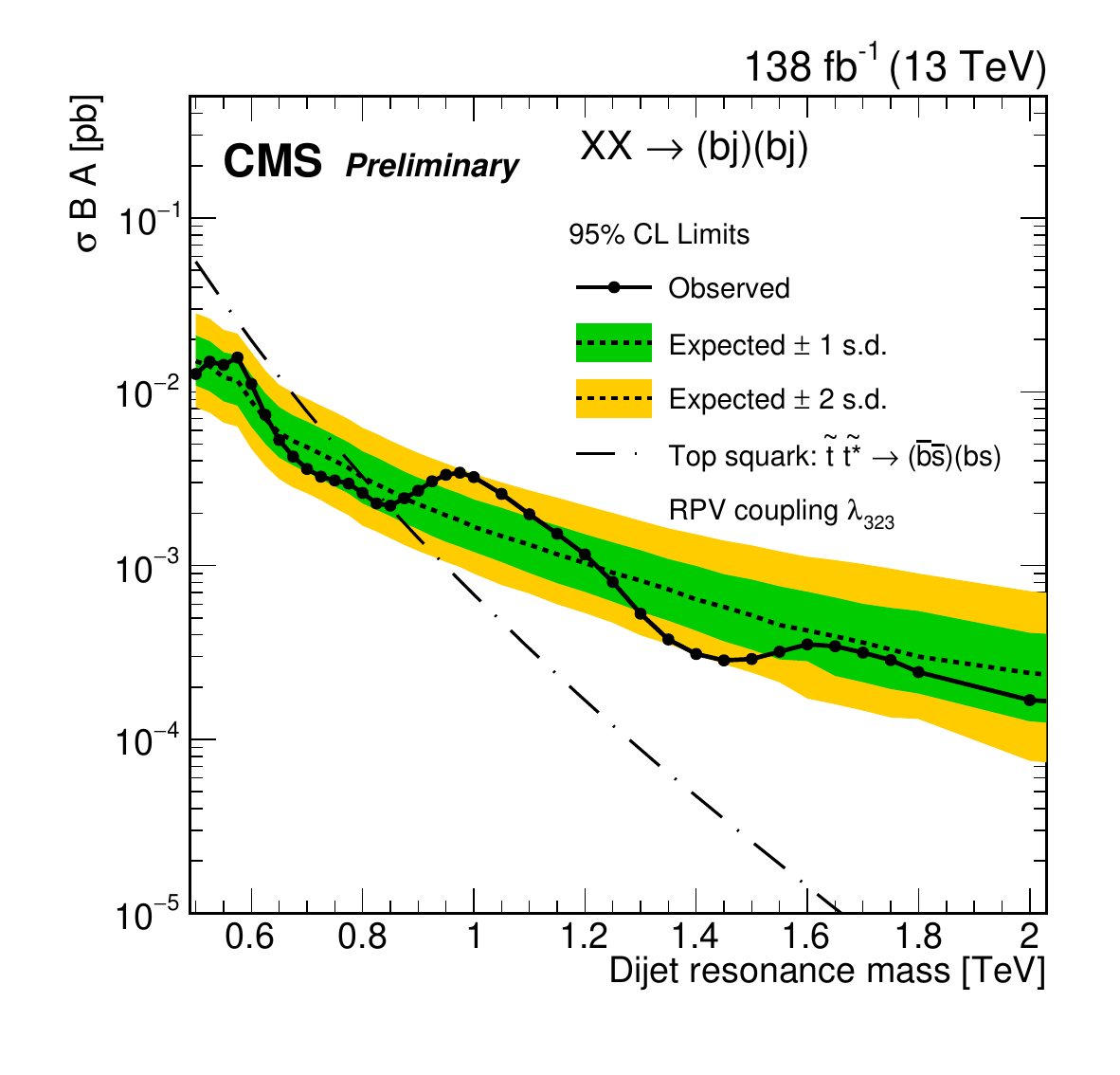}
\includegraphics[width=0.49\linewidth]{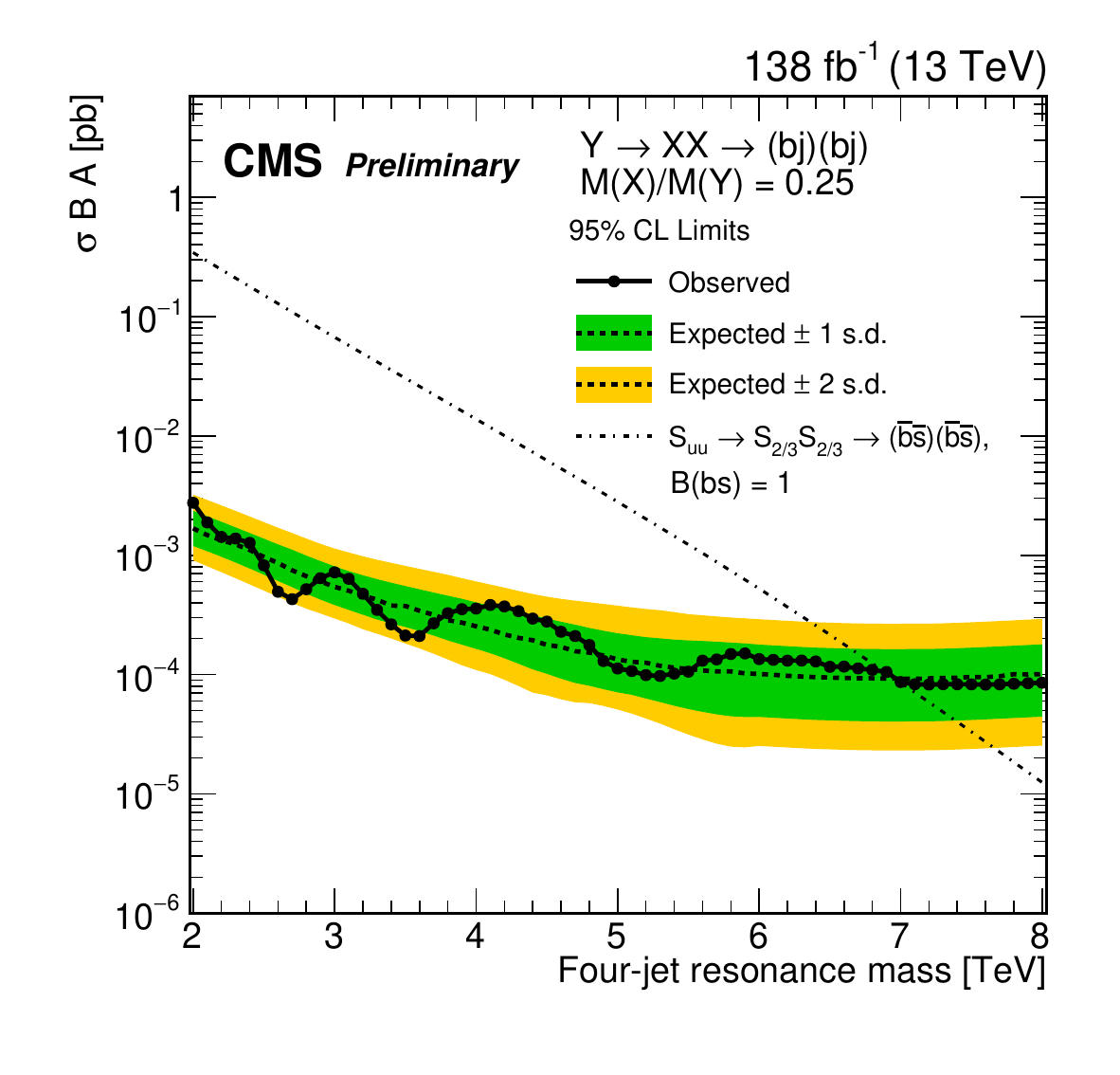}
\caption{Expected and observed 95\% CL upper limits for the nonresonant RPV top-squark search (left)
and the resonant diquark-to-diquarks search (right).
Limits are compared to the predicted cross-section of the RPV model and the diquark-to-diquarks model, respectively.}
\label{fig:pairdijet_b}
\end{figure}

No significant evidence for new resonances is observed.
In the nonresonant interpretation, top squarks are excluded for masses between 0.5 and 0.85~TeV.
In the resonant interpretation, heavy diquark states decaying to pairs of lighter diquarks
are excluded for masses between 2 and 7~TeV for nearly all mass ratios considered,
yielding the first LHC limits on resonant paired dijets with $b$ jets.
The exclusion upper limits for the two scenarios are reported in Figure~\ref{fig:pairdijet_b}.

\section{Conclusions}
The CMS high-mass search program continues to probe a wide range of resonant
and nonresonant signatures.
The Run~2 heavy-vector-boson combination shows that previously reported
local fluctuations are reduced when all major channels are analyzed together,
while the Run~3 $W^\prime \to \ell \nu$ result extends the direct sensitivity
to charged spin-1 resonances to nearly 6~TeV.
In dijet final states, both angular observables and paired-dijet topologies remain
powerful tools to test compositeness, dark-sector mediators, and multi-jet resonance models.

Across all analyses discussed in this contribution,
no significant deviation from the SM expectation is observed.
The large Run~3 data set still to be explored, together with continued improvements
in reconstruction, triggering, and theory modeling, will further extend the reach
of the CMS experiment for new physics at the highest accessible mass scales.

\section*{References}
\bibliography{malara}

\end{document}